\documentclass[showpacs,preprintnumbers,amsmath,amssymb]{revtex4}
\newcommand{\be}{\begin{equation}}
\newcommand{\ee}{\end{equation}}
\newcommand{\bea}{\begin{eqnarray}}
\newcommand{\eea}{\end{eqnarray}}

\newcommand{\integer}{\relax{\rm I\kern-.18em N}}

\begin{document}

\title{Quantum deformation of the Dirac bracket}
\author{M. I. Krivoruchenko$^{1,2}$, A. A. Raduta$^{2,3,4}$, Amand Faessler$^{2}$ \\
%EndAName
$^{1}${\small Institute for Theoretical and Experimental Physics, B.
Cheremushkinskaya 25}, {\small 117259 Moscow, Russia}\\
{\small \ }$^{2}${\small Institut f\"{u}r Theoretische Physik, T\"{u}bingen
Universit\"{a}t, Auf der Morgenstelle 14}, {\small D-72076 T\"{u}bingen,
Germany}\\
{\small \ }$^{3}${\small Department of Theoretical Physics and Mathematics,
Bucharest University}, {\small POBox MG11, Bucharest, Romania} \\
{\small \ }$^{4}${\small Institute of Physics and Nuclear Engineering, PO
Box MG06, Bucharest, Romania}}

\begin{abstract}
The quantum deformation of the Poisson bracket is the Moyal bracket. We
construct quantum deformation of the Dirac bracket for systems which admit
global symplectic basis for constraint functions. Equivalently, it can be
considered as an extension of the Moyal bracket to second-class constraints
systems and to gauge-invariant systems which become second class when
gauge-fixing conditions are imposed.
\end{abstract}

\maketitle

%%%%%%%%%%%%%%%%%%%%%%%%%%%%%%%%%%%%%%%%%%%%%%%%%%%%%%%%%%%%%%%%%%%%%%%%

\section{Introduction}

\setcounter{equation}{0} 
%%%%%%%%%%%%%%%%%%%%%%%%%%%%%%%%%%%%%%%%%%%%%%%%%%%%%%%%%%%%%%%%%%%%%%%%

%{\bf 1. Introduction.} 
The association rules between real functions in the phase space of
unconstrained classical systems and Hermitian operators in the Hilbert space
of the corresponding quantum mechanical systems are discussed since a long
time \cite{WEYL,WIGNER,GROE,MOYAL,MEHTA}. The commonly used association rule
proposed by Weyl \cite{WEYL} consists in replacing products of the canonical
variables with symmetrized products of operators of the canonical variables.
The Wigner function \cite{WIGNER} is associated to the density matrix.

To the lowest order in the Planck's constant, the product $\frak{f}\frak{g}$
of operators associated to functions $f$ and $g$ in the phase space
corresponds to the pointwise product $fg$. A deformation of the pointwise
product which keeps the association rule to all orders in the Planck's
constant is constructed by Groenewold \cite{GROE}. This product is known as
the star-product.

The Moyal bracket \cite{MOYAL} appears as a skew-symmetric part of the
star-product. It defines the representation of a commutator $-i/\hbar [,]$
in the space of the functions. The Moyal bracket is antisymmetric, coincides
with the Poisson bracket to the lowest order in the Planck's constant,
satisfies the Jacoby identity, and keeps the association rule. The bracket
satisfying these properties is essentially unique \cite{MEHTA}. The Moyal
bracket governs the quantum evolution of systems in the phase space like the
Poisson bracket governs the classical evolution. A survey of the
star-product and the Moyal quantization can be found in \cite{BAYEN,ZACH}.

The quantum dynamics of unconstrained systems can therefore be formulated in
the phase space in terms of the Hamiltonian and Wigner functions, with the
pointwise product replaced by the star-product. The average values of
quantum observables can be computed by averaging the symbols of the
Hermitian operators over the Wigner function.

The specific feature of gauge theories is the occurrence of constraints
which restrict the phase space of gauge-invariant systems to a submanifold.
A systematic Hamiltonian approach to gauge theories and general constraint
systems and the corresponding operator quantization schemes were developed
by Dirac \cite{DIRAC}.

Schemes based on the path integral method have also been proposed and found
to be useful for quantization of gauge theories \cite{HENN,FASL} and
second-class constraints systems \cite{BATA,BATY}.

Gauge systems are quantized by imposing gauge-fixing conditions which
convert them into second-class constraints systems. Anomalous gauge theories 
\cite{ANOM,AFAD,FSHA1,FSHA2,JOOO}, the $O(n)$ non-linear sigma model \cite
{BANE94,HONG04,KFRF}, many-body theories involving collective and elementary
degrees of freedom \cite{MARSH,BLAIZ,YAMA} are second class from the start.

If constraint equations are solved, the system can be restricted to the
constraint submanifold and treated accordingly as an unconstrained system.
In many cases, however, it is not possible to solve constraint equations.
The method proposed by Dirac for classical second-class constraints systems
solves the evolution problem by applying the Dirac bracket to functions of
canonical variables in the unconstrained phase space. It allows thereby to
avoid typical complications connected to restriction of the systems to
constraint submanifolds.

In this paper, we discuss second-class constraints systems from the Moyal
quantization perspective. The main problem we focus attention to is the
construction of quantum deformation of the Dirac bracket able to govern the
evolution of quantum constraint systems in the unconstrained phase space.
The guiding idea in development of the quantization scheme is that many
functions in the unconstrained phase space correspond to one and the same
physical observable. We thus come naturally to the notion of equivalence
classes of functions, and also operators. In the next Sect., we specify
equivalence classes of real functions in the unconstrained phase space and,
in Sect. III, of Hermitian operators in the Hilbert space. An association
rule between equivalence classes of operators and functions, based on the
Weyl's association rule, is established. In Appendix A, some useful
properties are derived for the skew-gradient projection in terms of which
the equivalence classes of functions and operators are defined. In Sect. IV,
we give a summary and, in Appendix B, proofs of the basic properties of the
Weyl's association rule and of the star-product. In Sect. V, the quantum
deformation of the Dirac bracket is constructed for systems which admit
global symplectic basis for constraint functions. In Appendix C, we provide
an explicit form of the lowest order $O(\hbar ^{2})$ quantum correction to
the Dirac bracket. As an application, in Sect. VI, the quantum deformation
of the Dirac bracket is used to formulate, in the unconstrained phase space,
an evolution equation for the Wigner function of an $n-1$-dimensional
spherical pendulum which represents a mechanical counterpart of the $O(n)$
non-linear sigma model.

%%%%%%%%%%%%%%%%%%%%%%%%%%%%%%%%%%%%%%%%%%%%%%%%%%%%%%%%%%%%%%%%%%%%%%%%

\section{Classical second-class constraints systems in the phase space}

\setcounter{equation}{0} 
%%%%%%%%%%%%%%%%%%%%%%%%%%%%%%%%%%%%%%%%%%%%%%%%%%%%%%%%%%%%%%%%%%%%%%%%

%{\bf 2. Second-class constraints systems in the phase space.} 
Second-class constraints $\mathcal{G}_{a} = 0$ with $a = 1,...,2m$ in an $2n$%
-dimensional unconstrained phase space $\xi
^{i}=(\phi^{1},...,\phi^{n},\pi^{1},...,\pi^{n})$ have the Poisson bracket
relations which form a non-degenerate $2m\times 2m$ matrix 
\begin{equation}
\det\{\mathcal{G}_{a},\mathcal{G}_{b}\}\ne 0.  \label{NONGEN}
\end{equation}

Two sets of the constraint functions are equivalent if they describe the
same constraint submanifold. One can make therefore non-degenerate
transformations on the constraint functions without changing the dynamics.

For an arbitrary given point of the constraint submanifold, there is a
neighborhood where one may find the equivalent constraint functions in terms
of which the Poisson bracket relations look like 
\begin{equation}
\{\mathcal{G}_{a},\mathcal{G}_{b}\}={}\mathcal{I}_{ab}  \label{SB}
\end{equation}
where 
\begin{equation}
\mathcal{I}_{ab}=\left\| 
\begin{array}{ll}
0 & E_{m} \\ 
-E_{m} & 0
\end{array}
\right\| ,  \label{SMAT}
\end{equation}
with $E_{m}$ being the unity $m\times m$ matrix, $\mathcal{I}_{ab}\mathcal{I}%
_{bc}=-\delta _{ac}$. The upper and lower indices of vectors are
discriminated according to the rules $\mathcal{T}_{a}=\mathcal{I}_{ab}%
\mathcal{T}^{b}$, $\mathcal{T}^{a}=\mathcal{I}^{ab}\mathcal{T}_{b}$, $%
\mathcal{I}^{ab}=-\mathcal{I}_{ab}$, $\{{}\mathcal{G}^{a},{}\mathcal{G}%
_{b}\}=\delta _{b}^{a}$, etc. The scalar product $\mathcal{T}^{a}\mathcal{Y}%
_{a}$ is invariant with respect to the group of linear symplectic
transformations $Sp(2m)$.

The global symplectic basis (\ref{SB}) for constraint functions exists
obviously for $m=1$ and, also, for systems of point particles under
second-class holonomic constraints \cite{KFRF} and second-class
non-holonomic constraints satisfying the Frobenius' condition \cite
{Krivoruchenko:2005wb}. The global existence of the basis (\ref{SB}) is
proved for systems with one primary constraint \cite{MITRA} and for a
broader set of systems using additional assumptions \cite{MITRA,VYTHE1}. In
general case, the global existence of the symplectic basis for constraint
functions is an opened question.

The basis (\ref{SB}) always exists locally, i.e., in a finite neghborhood of
any point of the constraint submanifold \cite
{Krivoruchenko:2005wb,MITRA,VYTHE1}. This is sufficient for needs of the
perturbation theory. The formalism presented in this work can therefore to
be used to fomulate, in the sense of the perturbation theory, the evolution
problem of any second-class constraints system in the unconstrained phase
space. Eqs.(\ref{SB}) follow from the local existence of the standard
canonical coordinate system \cite{MASKAWA} where ${}\mathcal{G}_{a}$ with $%
a=1,...,m$ play the role of the first canonical coordinates and ${}\mathcal{G%
}_{a}$ with $a=m+1,...,2m$ play the role of the first canonical momenta.

Let us construct skew-gradient projections $\xi _{s}(\xi )$ of the canonical
variables $\xi $ onto the constraint submanifold $\mathcal{G}_{a}(\xi )=0$
using phase flows generated by the constraint functions. From equations 
\begin{equation}
\{\xi _{s}(\xi ),\mathcal{G}_{a}(\xi )\}=0  \label{CG}
\end{equation}
using the symplectic basis (\ref{SB}) for the constraints and expanding 
\begin{equation}
\xi _{s}(\xi )=\xi +X^{a}\mathcal{G}_{a}+\frac{1}{2}X^{ab}\mathcal{G}_{a}%
\mathcal{G}_{b}+...
\end{equation}
in the power series of $\mathcal{G}_{a}$, one gets 
\begin{equation}
\xi _{s}(\xi )=\sum_{k=0}^{\infty }\frac{1}{k!}\{...\{\{\xi ,\mathcal{G}%
^{a_{1}}\},\mathcal{G}^{a_{2}}\},...\mathcal{G}^{a_{k}}\}\mathcal{G}_{a_{1}}%
\mathcal{G}_{a_{2}}...\mathcal{G}_{a_{k}}.  \label{SGRAD}
\end{equation}
One can show (see Appendix A) that any function $f(\xi )$ projected onto the
constraint submanifold 
\begin{equation}
f_{s}(\xi )=\sum_{k=0}^{\infty }\frac{1}{k!}\{...\{\{f(\xi ),\mathcal{G}%
^{a_{1}}\},\mathcal{G}^{a_{2}}\},...\mathcal{G}^{a_{k}}\}\mathcal{G}_{a_{1}}%
\mathcal{G}_{a_{2}}...\mathcal{G}_{a_{k}}  \label{FSG}
\end{equation}
satisfies 
\begin{equation}
f_{s}(\xi )=f(\xi _{s}(\xi )).  \label{FSSF}
\end{equation}

Eq.(\ref{CG}) tells us that variations of $\xi _{s}(\xi )$ along the phase
flows generated by the constraint functions $\mathcal{G}_{a}(\xi )$ are
zero. It means that $\xi _{s}(\xi )$ belongs to the constraint submanifold 
\begin{equation}
\mathcal{G}_{a}(\xi _{s}(\xi ))=0.  \label{on}
\end{equation}

For any function $f(\xi )$, one gets $\{\mathcal{G}_{a}(\xi ),f(\xi _{s}(\xi
))\}=0.$ The reciprocal statement is also true: The coordinates on the
constraint submanifold can be parameterized by $\xi _{s}$. The coordinates
describing shifts from the constraint submanifold can be parameterized by $%
\mathcal{G}_{a}$. The functions $f$ can be presented by $f=f(\xi _{s},%
\mathcal{G}_{a})$. If $f$ is identically in involution with $\mathcal{G}_{a}$%
, it depends on $\xi _{s}$ only. This can be summarized by 
\begin{equation}
\{\mathcal{G}_{a},f\}=0\leftrightarrow f=f(\xi _{s}(\xi )).  \label{XO}
\end{equation}

Eqs.(\ref{on}) and (\ref{XO}) become selfevident if one works in the
standard canonical coordinate system.

An average of a function $f(\xi)$ is calculated using the probability
density distribution $\rho(\xi)$ and the Liouville measure restricted to the
constraint submanifold \cite{FADD}: 
\begin{equation}
<f>=\int \frac{d^{2n}\xi}{(2\pi)^{n}} (2\pi)^{m}\prod_{a=1}^{2m} \delta (%
\mathcal{G}_{a}(\xi)) f(\xi)\rho(\xi).  \label{FAMEASURE}
\end{equation}
On the constraint submanifold $\xi _{s}(\xi ) = \xi$, so $f(\xi )$ and $%
\rho(\xi)$ can be replaced with $f_{s}(\xi )$ and $\rho_{s}(\xi )$.

There exist therefore equivalence classes of functions \cite{WI} in the
unconstrained phase space: 
\begin{equation}
f(\xi )\sim g(\xi )\leftrightarrow f_{s}(\xi )=g_{s}(\xi ).  \label{EQ}
\end{equation}
Here, $f(\xi )\sim g(\xi )$ means that the functions are equal in the weak
sense, $f(\xi )\approx g(\xi )$, i.e., on the constraint submanifold. We
shall see that the symbols $\sim $ and $\approx $ acquire distinct meaning
upon quantization. Note that $f(\xi )\sim f_{s}(\xi ).$ Eqs.(\ref{FSSF}) and
(\ref{on}) imply $\mathcal{G}_{a}\sim 0.$

Given the hamiltonian function ${}\mathcal{H}$, the evolution of a function $%
f$ is described using the Dirac bracket 
\begin{equation}
\frac{\partial }{\partial t}f=\{f,{}\mathcal{H}\}_{D}.  \label{EV}
\end{equation}
In the symplectic basis (\ref{SB}), the Dirac bracket looks like 
\begin{equation}
\{f,g\}_{D}=\{f,g\}+\{f,\mathcal{G}^{a}\}\{\mathcal{G}_{a},g\}.  \label{DBNA}
\end{equation}
On the constraint submanifold, one has 
\begin{equation}
\{f,g\}_{D}=\{f,g_{s}\}=\{f_{s},g\}=\{f_{s},g_{s}\}.  \label{DB1}
\end{equation}

Two hamiltonian functions are equivalent if they generate within the
constraint submanifold identical phase flows. The components of the
hamiltonian phase flow, which belong to a subspace spanned at the constraint
submanifold by phase flows of the constraint functions, do not affect the
dynamics and could be different. $\mathcal{H}{}$ and ${}\mathcal{H}_{s}$ are
thereby equivalent, so Eq.(\ref{EQ}) characterizes an equivalence class for
the hamiltonian functions either. Among functions of this class, ${}\mathcal{%
H}_{s}$ is the one whose phase flow is skew-orthogonal to phase flows of the
constraint functions.

Replacing ${}\mathcal{H}$ with ${}\mathcal{H}_{s}$, one can rewrite the
evolution equation in terms of the Poisson bracket (cf. Eq.(\ref{EV})): 
\begin{equation}
\frac{\partial }{\partial t}f=\{f,{}\mathcal{H}_{s}\}.  \label{PEV}
\end{equation}

Eq.(\ref{PEV}) applied to $f(t)$ and $g(t)$ which belong at $t=0$ to the
same equivalence class, provides with the use of Eq.(\ref{A90}) and the
initial condition $f_{s}(0)=g_{s}(0)$, $f_{s}(t)=g_{s}(t)$. The equivalence
relations (\ref{EQ}) are therefore preserved during the evolution.

If we denote the equivalence class of a function $f(\xi )$ as $\mathcal{E}%
_{f}$, the sum of two equivalence classes $\mathcal{E}_{f}$ and $%
\mathcal{E}_{g}$ can be defined as $\mathcal{E}_{f}+\mathcal{E}_{g}=\mathcal{%
E}_{f+g}$ on line with $f_{s}+g_{s}=(f+g)_{s}$, the associative product can
be identified with $\mathcal{E}_{fg}$, while the skew-symmetric Dirac
bracket can be defined as $\{\mathcal{E}_{f},\mathcal{E}_{g}\}_{D}=\mathcal{E%
}_{\{f,g\}_{D}}$. These operations satisfy the Leibniz' law, 
\begin{equation}
\{\mathcal{E}_{f}\mathcal{E}_{g},\mathcal{E}_{h}\}_{D}=\mathcal{E}_{f}\{%
\mathcal{E}_{g},\mathcal{E}_{h}\}_{D}+\{\mathcal{E}_{f},\mathcal{E}_{h}\}_{D}%
\mathcal{E}_{g},
\end{equation}
and the Jacoby identity, 
\begin{equation}
\{\{\mathcal{E}_{f},\mathcal{E}_{g}\}_{D},\mathcal{E}_{h}\}_{D}+\{\{\mathcal{%
E}_{g},\mathcal{E}_{h}\}_{D},\mathcal{E}_{f},\}_{D}+\{\{\mathcal{E}_{h},%
\mathcal{E}_{f}\}_{D},\mathcal{E}_{g}\}_{D}=0.
\end{equation}
The associative product does not depend, in virtue of Eq.(\ref{MULTPR}), on
the choice of representatives of the equivalence classes. The Dirac bracket
can be calculated for arbitrary representatives of the equivalence classes
either. Indeed, $f\sim g$ implies 
\[
\{f,h\}_{D}\sim
(\{f,h\}_{D})_{s}=\{f_{s},h_{s}\}_{D}=\{g_{s},h_{s}\}_{D}=(\{g,h\}_{D})_{s}%
\sim \{g,h\}_{D}, 
\]
where the use is made of Eq.(\ref{A2}).

\textit{The physical observables in second-class constraints systems are
associated with the equivalence classes of real functions in the
unconstrained phase space. The equivalence classes }$\mathcal{E}_{f}$\textit{%
\ constitute a vector space }$\mathcal{O}$\textit{${}$ equipped with two
multiplication operations, the associative pointwise product and the
skew-symmetric Dirac bracket $\{,\}_{D},$ which confer ${}$}$\mathcal{O}$%
\textit{\ a Poisson algebra structure. }

The one-to-one mapping $\mathcal{E}_{f}\leftrightarrow f_{s}$ induces a
Poisson algebra structure on the vector space of projected functions$.$ The
sum $\mathcal{E}_{f}+\mathcal{E}_{g}$ converts to $f_{s}+g_{s}$, the
associative product $\mathcal{E}_{f}\mathcal{E}_{g}$ converts to the
pointwise product $f_{s}g_{s}$, while the Dirac bracket $\{\mathcal{E}_{f},%
\mathcal{E}_{g}\}_{D}$ becomes the Poisson bracket (cf. transition from (\ref
{EV}) to (\ref{PEV})): 
\begin{equation}
\{f_{s},g_{s}\}_{D}=\{f_{s},g_{s}\}.  \label{DBasPB}
\end{equation}
These operations satisfy the Leibniz' law and the Jacoby identity and, since 
$(f_{s}+g_{s})_{s}=f_{s}+g_{s}$, $(f_{s}g_{s})_{s}=f_{s}g_{s}$, and $%
\{f_{s},h_{s}\}_{s}=\{f_{s},h_{s}\}$ (cf. Eqs.(\ref{MULTPR}) and (\ref{A90}%
)), keep the vector space of projected functions closed.

%%%%%%%%%%%%%%%%%%%%%%%%%%%%%%%%%%%%%%%%%%%%%%%%%%%%%%%%%%%%%%%%%%%%%%%%

\section{Quantum second-class constraints systems in the Hilbert space}

\setcounter{equation}{0} 
%%%%%%%%%%%%%%%%%%%%%%%%%%%%%%%%%%%%%%%%%%%%%%%%%%%%%%%%%%%%%%%%%%%%%%%%

%{\bf 3. Second-class constraints systems in the Hilbert space.}
The systems are quantized by the algebra mapping $\xi ^{i}\to \frak{x}^{i}$
and $\{,\}\to -i/{\hbar }[,]$. To any function $f(\xi )$ in the
unconstrained phase space one may associate an operator $\frak{f}$ in the
corresponding Hilbert space. In particular the hamiltonian function $%
\mathcal{H}(\xi )$ and the constraint functions $\mathcal{G}_{a}(\xi )$
correspond to the operators $\frak{H}$ and $\frak{G}_{a}$, respectively. By
this mapping the quantal image is called operator. The reverse mapping
associates to an operator $\frak{f}$ a symbol $f(\xi )$.

Eqs.(\ref{SB}) become 
\begin{equation}
[\frak{G}_{a},\frak{G}_{b}] = i{\hbar}\mathcal{I}_{ab}.  \label{QSB}
\end{equation}
The Weyl's association rule applied to Eqs.(\ref{SB}) yields Eqs.(\ref{QSB}),
provided the quantization is performed in the standard canonical coordinate
system.

Any operator in the Hilbert space can be represented as a function of $2n$
operators $\frak{x}^{i}$ associated to $2n$ canonical variables $\xi^{i}$.

Let us construct $2n$ operators $\frak{x}_{s}^{i}$ associated to the
projected variables (\ref{SGRAD}). They commute with the constraints 
\begin{equation}
[\frak{x}_{s},\frak{G}_{a}]=0.  \label{QCG}
\end{equation}
The analogue of Eq.(\ref{SGRAD}) looks like 
\begin{equation}
\frak{x} _{s}=\sum_{k=0}^{\infty }\frac{(-i/{\hbar})^{k}}{k!}[...[[\frak{x} , 
\frak{G}^{a_{1}}],\frak{G}^{a_{2}}],...\frak{G}^{a_{k}}] \frak{G}_{a_{1}}%
\frak{G}_{a_{2}}...\frak{G}_{a_{k}}.  \label{QSGRAD}
\end{equation}

Among the operators acting in the Hilbert space, one expects that,
equivalence classes exist. For an arbitrary operator $\frak{f}$, a projected
operator $\frak{f}_{s}$ can be constructed as follows: 
\begin{equation}
\frak{f}_{s}=\sum_{k=0}^{\infty }\frac{(-i/{\hbar })^{k}}{k!}[...[[\frak{f},%
\frak{G}^{a_{1}}],\frak{G}^{a_{2}}],...\frak{G}^{a_{k}}]\frak{G}_{a_{1}}%
\frak{G}_{a_{2}}...\frak{G}_{a_{k}}.  \label{QFSG}
\end{equation}
One has 
\begin{equation}
\lbrack \frak{f}_{s},\frak{G}_{a}]=0  \label{QQCG}
\end{equation}
and 
\begin{equation}
(\frak{f}\frak{g}_{s})_{s}=(\frak{f}_{s}\frak{g})_{s}=\frak{f}_{s}\frak{g}%
_{s}.  \label{POWE}
\end{equation}

Two operators $\frak{f}$ and $\frak{g}$ belong to the same equivalence class
provided $\frak{f}_{s}=\frak{g}_{s}$, i.e., 
\begin{equation}
\frak{f}\sim \frak{g}\leftrightarrow \frak{f}_{s}=\frak{g}_{s}.  \label{QEQ}
\end{equation}

The Dirac's quantization method of second-class constraints systems \cite
{DIRAC} consists in constructing operators reproducing the Dirac bracket for
canonical variables and taking constraints to be operator equations. In the
classical limit, the commutators of operators (\ref{QSGRAD}) satisfy the
Dirac bracket relations for canonical variables on the constraint
submanifold. Furthermore, $(\frak{G}^{a})_{s}=0$, and so $\frak{G}^{a} \sim 0$.

As a consequence of the Jacoby identity and Eq.(\ref{QSB}), the operator 
\begin{equation}
\frak{C}^{a_{1}...a_{k}}=[...[[{\ },\frak{G}^{a_{1}}],\frak{G}^{a_{2}}],...%
\frak{G}^{a_{k}}]  \label{ssss}
\end{equation}
entering Eqs.(\ref{QSGRAD}) and (\ref{QFSG}) is symmetric with respect to
permutations of $a_{1},...,a_{k}$. Any contraction of the upper indices with 
$\mathcal{I}_{ab}$ annihilates (\ref{ssss}). It follows that 
\begin{equation}
\lbrack \frak{C}^{a_{1}...a_{i}...a_{k}},\frak{G}_{a_{i}}]=0.  \label{dddd}
\end{equation}
The position of the constraint operators with the lower indices and of the
operator (\ref{ssss}) in Eqs.(\ref{QSGRAD}) and (\ref{QFSG}) is not
important. One can place, e.g., $\frak{G}_{a_{1}}$ on the first position, $%
\frak{C}^{a_{1}...a_{k}}$ on the second position, etc.

In order to calculate the average value of an operator, one has to construct
the quantal image of the delta functions product entering Eq.(\ref{FAMEASURE}%
). The projection operator can be written as follows 
\begin{equation}
\frak{P} = \int { \frac{d^{2m}\lambda}{(2\pi \hbar)^{m}} } \prod_{a=1}^{2m}\exp( \frac{i}{\hbar } \frak{G}^{a}
\lambda_{a}).  \label{QQQEV}
\end{equation}
In the classical limit, one recovers the product of the delta functions.

Let us chose a basis in the Hilbert space in which the first $m$ constraint
operators are diagonal, 
\begin{equation}
\frak{G}^{a}|g,g_{*}>=g^{a}|g,g_{*}>,  \label{A1}
\end{equation}
for $a=1,...,m$. Congruous to this equation, $\frak{G}^{a}$ might be taken 
as momentum operators. The last $m$ constraint operators can be treated as
quantal coordinates. The additional $n-m$ eigenvalues are denoted by $g_{*}$.

The projection operator $\frak{P}$ acts as follows: 
\begin{equation}
\frak{P} |g,g_{*}>  = |0,g_{*}>.  \label{A7}
\end{equation}
To arrive at this equation, we split $\frak{G}^{a} = (\mathbf{P}^{A},-\mathbf{Q}^{A})$ and 
$\lambda^{a} = (\lambda_{A}^{\prime}, \lambda_{A}^{\prime \prime})$ and write
\begin{eqnarray*}
\frak{P}|g,g_{*} &>&=\int \frac{d^{2m}\lambda }{(2\pi \hbar )^{m}}%
\prod_{a=1}^{2m}\exp (\frac{i}{\hbar }\frak{G}^{a}\lambda _{a})
                               |g,g_{*}>
\\
&=&\int d^{m}\lambda^{\prime}\frac{d^{m}\lambda^{\prime \prime}}{(2\pi \hbar )^{m}} 
\prod_{A=1}^{m}\exp (  \frac{i}{\hbar }\mathbf{P}^{A}\lambda_{A}^{ \prime        })
\prod_{B=1}^{m}\exp (- \frac{i}{\hbar }\mathbf{Q}^{B}\lambda_{B}^{ \prime \prime })
                               |g,g_{*}> \\
&=&\int d^{m}\lambda^{\prime}\frac{d^{m}\lambda^{\prime \prime}}{(2\pi \hbar )^{m}} 
\prod_{A=1}^{m}\exp (  \frac{i}{\hbar }\mathbf{P}^{A}\lambda_{A  }^{\prime })
                               |g-\lambda^{\prime \prime},g_{*}> \\
&=&\int d^{m}\lambda^{\prime}\frac{d^{m}\lambda^{\prime \prime}}{(2\pi \hbar )^{m}} 
\prod_{A=1}^{m}\exp (  \frac{i}{\hbar }(g^{A}-\lambda_{A}^{\prime \prime })\lambda _{A}^{\prime })
                               |g-\lambda^{\prime \prime},g_{*}> \\
&=&\int d^{m}\lambda ^{\prime \prime }\prod_{A=1}^{m}\delta (g^{A}-\lambda_{A}^{\prime \prime })
                               |g-\lambda^{\prime \prime },g_{*}> \\
&=&|0,g_{*}>
\end{eqnarray*}
The average value of an operator $\frak{f}$, 
\begin{equation}
<\frak{f}>=Tr[\frak{P}\frak{f}_{s}\frak{r}_{s}]  \label{QQEV}
\end{equation}
where $\frak{r}=\frak{r}^{+}$ is the density matrix, can be transformed to
give 
\begin{eqnarray}
Tr[\frak{P}\frak{f}_{s}\frak{r}_{s}] &=&\int { \frac{d^{m}g d^{n-m}g_{*}}{(2\pi \hbar)^{n}} 
<g,g_{*}| \frak{P}\frak{f}_{s}\frak{r}_{s}|g,g_{*}>}  \nonumber \\
&=&\int {  \frac{d^{m}g d^{n-m}g_{*}}{(2\pi \hbar)^{n}}  <g,g_{*}|\frak{f}_{s}\frak{r}_{s}|0,g_{*}>}
 = \int { \frac{d^{n-m}g_{*}}{(2\pi \hbar)^{n-m}} <0,g_{*}|\frak{f}_{s}\frak{r}_{s}|0,g_{*}>}.  \label{A3}
\end{eqnarray}
The average values are determined by the physical subspace of the Hilbert
space, spanned by the vectors $|0,g_{*}>$.

These vectors satisfy the equation 
\begin{equation}
\frak{G}^{a}|0,g_{*}>=0  \label{DSC}
\end{equation}
which can be recognized as the Dirac's supplementary condition \cite{DIRAC}
of an equivalent gauge system \cite{KFRF,Krivoruchenko:2005wb,MITRA,VYTHE1},
where $\frak{G}^{a}$ with $a=1,...,m$ are gauge generators and $\frak{G}^{a}$
with $a=m+1,...,2m$ are gauge-fixing operators.

All density matrices from a given equivalence class correspond to a single
physical state, while operators from the same equivalence class have equal
average values.

The quantum evolution equation can be written as an extension of the
classical evolution equation (\ref{PEV}) 
\begin{equation}
i{\hbar }\frac{d}{dt}\frak{f}=[\frak{f},\frak{H}_{s}]  \label{QEV}
\end{equation}
where $\frak{H}_{s}$ is the projection (\ref{QFSG}) of the Hamiltonian $%
\frak{H}$. The evolution has the property that at any time $\frak{f}(t)\sim 
\frak{g}(t)$ if $\frak{f}(0)\sim \frak{g}(0)$. This is suggested by the
equation 
\begin{equation}
\lbrack \frak{f},\frak{g}_{s}]_{s}=[\frak{f}_{s},\frak{g}_{s}]
\label{PROJCOM}
\end{equation}
which is evident due to $(\frak{f}+\frak{g})_{s}=\frak{f}_{s}+\frak{g}_{s}$
and Eq.(\ref{POWE}).

%%%%%%%%%%%%%%%%%%%%%%%%%%%%%%%%%%%%%%%%%%%%%%%%%%%%%%%%%%%%%%%%%%%%%%%%
\section{Weyl's association rule and the star-product}

\setcounter{equation}{0} 
%%%%%%%%%%%%%%%%%%%%%%%%%%%%%%%%%%%%%%%%%%%%%%%%%%%%%%%%%%%%%%%%%%%%%%%%

%{\bf 4. Quantum deformation of the phase space.} 
The Weyl's association rule can be formulated in a comprehensive way in
terms of the operator function \cite{BERE} 
\begin{equation}
\frak{\tilde{B}}(\eta )=\exp (\frac{i}{\hbar }\eta _{k}\frak{x}^{k})
\label{PT}
\end{equation}
whose Fourier transform, 
\begin{equation}
\frak{B}(\xi )=\int \frac{d^{2n}\eta }{(2\pi \hbar )^{n}}\exp (-\frac{i}{%
\hbar }\eta _{k}\xi ^{k})\frak{\tilde{B}}(\eta ),  \label{P}
\end{equation}
has the properties (see Appendix B) 
\begin{eqnarray}
\frak{B}(\xi )^{+} &=&\frak{B}(\xi ),  \label{PROP} \\
Tr[\frak{B}(\xi )] &=&1,  \label{PROP2} \\
\int \frac{d^{2n}\xi }{(2\pi \hbar )^{n}}\frak{B}(\xi ) &=&\frak{1},
\label{PROP3} \\
\int \frac{d^{2n}\xi }{(2\pi \hbar )^{n}}\frak{B}(\xi )Tr[\frak{B}(\xi )%
\frak{f}] &=&\frak{f},  \label{PROP4} \\
Tr[\frak{B}(\xi )\frak{B}(\xi ^{\prime })] &=&(2\pi \hbar )^{n}\delta
^{2n}(\xi -\xi ^{\prime }),  \label{PROP5} \\
\frak{B}(\xi )\exp (-\frac{i\hbar }{2}\mathcal{P}_{\xi \xi ^{\prime }})\frak{%
B}(\xi ^{\prime }) &=&(2\pi \hbar )^{n}\delta ^{2n}(\xi -\xi ^{\prime })%
\frak{B}(\xi ^{\prime }).  \label{PROP6}
\end{eqnarray}
Here, 
\[
\mathcal{P}_{\xi \xi ^{\prime }}\mathcal{=}-I{}{}^{kl}\overleftarrow{\frac{%
\partial }{\partial \xi ^{k}}}\overrightarrow{\frac{\partial }{\partial \xi
^{\prime l}}} 
\]
is the so-called Poisson operator. The matrix ${}I^{kl}$ looks similarly to
the matrix (\ref{SMAT}) 
\begin{equation}
I{}^{kl}=\left\| 
\begin{array}{ll}
0 & -E_{n} \\ 
E_{n} & 0
\end{array}
\right\| ,
\end{equation}
with $E_{n}$ being the $n\times n$ identity matrix.

Equations 
\begin{eqnarray}
f(\xi ) &=&Tr[\frak{B}(\xi )\frak{f}],  \label{S} \\
\frak{f} &=&\int \frac{d^{2n}\xi }{(2\pi \hbar )^{n}}f(\xi )\frak{B}(\xi ).
\label{INV}
\end{eqnarray}
define the Weyl's association rule.

The canonical variables appear as symbols of operators of the coordinates
and momenta: $\xi ^{i}=Tr[\frak{B}(\xi )\frak{x}^{i}]$.

The phase space of quantum systems is equipped with the Groenewold
star-product \cite{GROE}. Given two functions 
\begin{eqnarray*}
f(\xi ) &=&Tr[\frak{B}(\xi )\frak{f}], \\
g(\xi ) &=&Tr[\frak{B}(\xi )\frak{g}],
\end{eqnarray*}
one can construct a third function 
\begin{equation}
f(\xi )\star g(\xi )=Tr[\frak{B}(\xi )\frak{fg}].  \label{GR}
\end{equation}

%%%%%%%%%%%%%%%%%%%%%%%%%%%%%%%%%%%%%%%%%%%%%%%%%%%%%%%%%%%%%%

The star-product is an associative operation. It splits into symmetric and
antisymmetric parts: 
\begin{equation}
f\star g=f\circ g+\frac{i\hbar }{2}f\wedge g.  \label{SSM}
\end{equation}
The explicit form of the star-product can be derived from Eqs.(\ref{PROP}) - (\ref{PROP6}): 
\begin{equation}
f(\xi )\star g(\xi )=f(\xi )\exp (\frac{i\hbar }{2}\mathcal{P})g(\xi ),
\label{EG}
\end{equation}
where $\mathcal{P}=\mathcal{P}_{\xi \xi }$ and therefore 
\begin{eqnarray}
f(\xi )\circ g(\xi ) &=&f(\xi )\cos (\frac{\hbar }{2}\mathcal{P})g(\xi ),
\label{ES} \\
f(\xi )\wedge g(\xi ) &=&f(\xi )\frac{2}{\hbar }\sin (\frac{\hbar }{2}%
\mathcal{P})g(\xi ).  \label{EA}
\end{eqnarray}
The Planck's constant $\hbar $ appears as a quantum deformation parameter.
The antisymmetric product $f(\xi )\wedge g(\xi )$ is known under the name of
Moyal bracket. The classical limit of the Moyal bracket is the Poisson
bracket: 
\begin{equation}
\lim_{\hbar \rightarrow 0}f(\xi )\wedge g(\xi )=\{f(\xi ),g(\xi )\}.
\label{LIMIT}
\end{equation}

The star-product and the Moyal bracket obey the Leibniz' law 
\begin{equation}
f\wedge (g\star h)=(f\wedge g)\star h+g\star (f\wedge h).  \label{LEIB}
\end{equation}
This equation is valid separately for symmetric and antisymmetric parts of
the star-product. In the last case, Eq.(\ref{LEIB}) provides the Jacoby
identity.

%%%%%%%%%%%%%%%%%%%%%%%%%%%%%%%%%%%%%%%%%%%%%%%%%%%%%%%%%%%%%%%%%%%%%%%%

\section{Quantum second-class constraints systems in the phase space}

\setcounter{equation}{0} 
%%%%%%%%%%%%%%%%%%%%%%%%%%%%%%%%%%%%%%%%%%%%%%%%%%%%%%%%%%%%%%%%%%%%%%%%

%{\bf 5. Second-class constraints systems in the deformed phase space.} 
The phase space of quantum systems is endowed with the star-product. The
corresponding Hamiltonian ${H}(\xi )$ and constraint functions ${G}_{a}(\xi
) $ appear as symbols of the operators $\frak{H}$ and $\frak{G}_{a}$,
respectively, as prescribed by Eq.(\ref{S}). If the quantization is done in
the standard canonical coordinate system, then ${G}_{a}(\xi )=\mathcal{G}%
{}_{a}(\xi )$. In general case, the equality holds in the classical limit,
i.e., 
\begin{equation}
\lim_{{\hbar }\rightarrow 0}{G}_{a}(\xi )=\mathcal{G}{}_{a}(\xi ).
\label{LG}
\end{equation}

Classical systems represent a limiting case of quantum systems, not vice
versa, so many quantum systems with the same classical limit usually exist.
Some ambiguities arise due to the so-called ''operator ordering problem''.
There exist, as a consequence, several association rules the most popular of
which is the Weyl's rule. The explicit forms of distinct association rules
are provided by Mehta \cite{MEHTA}. The second-class constraints systems
have additional ambiguities connected to the choice of constraint functions $%
{G}_{a}(\xi )$. Here, we require only that ${H}(\xi )$ and ${G}_{a}(\xi )$
exist and tend, respectively, to $\mathcal{H}(\xi )$ and $\mathcal{G}%
_{a}(\xi )$ as ${\hbar }\rightarrow 0$.

The analogue of Eqs.(\ref{SB}) and (\ref{QSB}), which specifies the
symplectic basis for the constraint functions ${G}_{a}$, has the form 
\begin{equation}
{G}_{a}(\xi )\wedge {G}_{b}(\xi )=\mathcal{I}{}_{ab}.  \label{SBAS}
\end{equation}
These equations are automatically fulfilled provided the constraint
operators $\frak{G}_{a}$ obey Eqs.(\ref{QSB}).

The skew-gradient projections, analogous to Eqs.(\ref{CG}) and (\ref{QCG}),
are defined by 
\begin{equation}
\xi _{t}(\xi )\wedge {G}_{a}(\xi )=0.  \label{CG3}
\end{equation}

The projected canonical variables have the form 
\begin{equation}
\xi _{t}(\xi )=\sum_{k=0}^{\infty }\frac{1}{k!}(...((\xi \wedge {G}%
^{a_{1}})\wedge {G}^{a_{2}})...\wedge {G}^{a_{k}})\circ {G}_{a_{1}}\circ {G}%
_{a_{2}}...\circ {G}_{a_{k}}  \label{SGRAD3}
\end{equation}
(cf. Eqs.(\ref{SGRAD}) and (\ref{QSGRAD})). The analogue of Eqs.(\ref{FSG})
and (\ref{QFSG}) is 
\begin{equation}
f_{t}(\xi )=\sum_{k=0}^{\infty }\frac{1}{k!}(...((f(\xi )\wedge {G}%
^{a_{1}})\wedge {G}^{a_{2}})...\wedge {G}^{a_{k}})\circ {G}_{a_{1}}\circ {G}%
_{a_{2}}...\circ {G}_{a_{k}}.  \label{SGRAD4}
\end{equation}
The operation $\circ $ is not associative in general. The order in which it
acts is, however, not important due to commutativity of the operators in
expressions (\ref{QSGRAD}) and (\ref{QFSG}). In the classical limit one has 
\begin{equation}
\lim_{\hbar \rightarrow 0}f_{t}=f_{s}.  \label{FT2FS}
\end{equation}

The terms $...\circ {G}_{a}$ entering Eq.(\ref{SGRAD3}) involve the
derivatives and therefore do not vanish when ${G}_{a}=0$, so the variables $%
\xi _{t}(\xi )\,$are at variances $O(\hbar ^{2})$ with the variables $\xi $
on the constraint submanifold. Respectively, $f_{t}(\xi )\ $and $f(\xi )\ $%
are at a variance $O(\hbar ^{2})$ on the constraint submanifold also.

The lowest order quantum corrections in Eq.(\ref{SGRAD4}) can originate from
the Moyal bracket inside of the largest group of terms inside of the round
brackets. In such a case, the symmetrized $\circ $-product can be replaced
with the desired accuracy by the pointwise product. On the constraint
submanifold, those terms do not contribute to the result and, respectively,
the Moyal bracket in Eq.(\ref{SGRAD4}) can be replaced with the Poisson
bracket. The second possible source of the quantum corrections is the
operation $\circ $ itself, as defined by Eq.(\ref{ES}). To the lowest order
in the Planck's constant, the fourth-order differential operator $\mathcal{P}%
^{2}$ appears$.$ The multiple product $(...)\circ {G}_{a_{1}}\circ {G}%
_{a_{2}}...\circ {G}_{a_{k}}$ can be calculated as $(...(((...)\circ {G}%
_{a_{1}})\circ {G}_{a_{2}})...)\circ {G}_{a_{k}}$. The zeroth and second
powers of $\hbar $ show up in the first three terms of the $k$-series only.
One gets, on the constraint submanifold, 
\begin{eqnarray}
f_{t}(\xi ) &=&f(\xi )-\frac{\hbar ^{2}}{8}\{f(\xi ),{G}^{a_{1}}\}\mathcal{P}%
^{2}{G}_{a_{1}}-\frac{\hbar ^{2}}{8}\frac{1}{2!}\{\{f(\xi ),{G}^{a_{1}}\},{G}%
^{a_{2}}\}{G}_{a_{1}}\mathcal{P}^{2}{G}_{a_{2}}  \nonumber \\
&&-\frac{\hbar ^{2}}{8}\frac{1}{3!}\{\{\{f(\xi ),{G}^{a_{1}}\},{G}^{a_{2}}\},%
{G}^{a_{3}}\}{G}_{a_{1}}{G}_{a_{2}}\mathcal{P}^{2}{G}_{a_{3}}+O(\hbar ^{4}).
\label{FT}
\end{eqnarray}
The operator $\mathcal{P}$ acts on the left and on the right as prescribed
by Eq.(\ref{ES}). The terms ${G}_{a}$ involving no $\mathcal{P}$ action
vanish on the constraint submanifold and are therefore omitted. The
projected functions differ in general from the original ones. The constraint
functions are, however, an exception, i.e., $({G}_{a})_{t}={0}$ on the
constraint submanifold and, furthermore, in the unconstrained phase space.

The equivalence relations between the operators Eq.(\ref{QEQ}) lead, under
the Weyl's association rule, to the equivalence relations between functions
in the phase space: 
\begin{equation}
f(\xi )\sim g(\xi )\leftrightarrow f_{t}(\xi )=g_{t}(\xi )  \label{EQRELA}
\end{equation}
The physical observables are mapped through the quantization procedure onto
the equivalence classes of functions. Since $f(\xi )\sim f_{t}(\xi )$ and $%
f(\xi )\neq f_{t}(\xi )$ on the constraint submanifold, the symbols $\sim $
and $\approx $ acquire distinct meaning.

The average value of a function $f(\xi)$ is defined as 
\begin{equation}
<f>=\int \frac{d^{2n}\xi}{(2\pi \hbar)^{n}} P(\xi)\star f_{t}(\xi)\star W_{t}(\xi)
\label{FAMEASURE3}
\end{equation}
where $P(\xi)$ is the symbol of the projection operator $\frak{P}$ and $W(\xi)$
is the Wigner function, i.e., the symbol of the density matrix $\frak{r}$.
In the classical limit 
\begin{eqnarray}
\lim_{\hbar \rightarrow 0}\frac{P(\xi)}{(2\pi \hbar)^{m}} = \prod_{a=1}^{2m} \delta (%
\mathcal{G}_{a}(\xi)).  \label{LIM}
\end{eqnarray}
If the quantization is done in the standard canonical coordinate system, $%
P(\xi)/(2\pi \hbar)^{m}$ turns to the product of the delta functions to all orders in $\hbar$.

The evolution equation which is analogue of Eqs.(\ref{PEV}) and (\ref{QEV}%
) takes the form 
\begin{equation}
\frac{\partial}{\partial t}f(\xi) = f(\xi) \wedge {H}_{t}(\xi).  \label{PEV2}
\end{equation}

The quantum deformation of the Dirac bracket represents a skew-symmetric
multiplication operation on the set of equivalence classes of functions in
the phase space of quantum systems.

Like in the classical case, the sum of two equivalence classes ${E}_{f}$ and 
${E}_{g}$ is defined by ${E}_{f}+{E}_{g}={E}_{f+g}$, the associative product 
${E}_{f}{E}_{g}$ is defined for quantum systems using the star-product as ${E%
}_{f_{t}\star g_{t}}$, and the skew-symmetric Dirac bracket is defined as $\{%
{E}_{f},{E}_{g}\}_{D}={E}_{f_{t}\wedge g_{t}}$. These operations satisfy the
Leibniz' law and the Jacoby identity. The associative product and the Dirac
bracket do not depend on the choice of representatives of the equivalence
classes. Since $({G}^{a})_{t}=0$, the deformed Dirac bracket vanishes for
any two equivalence classes the one of which contains a constraint function.

\textit{By quantization, the Dirac bracket is associated to the commutator
of two projected operators. The Moyal bracket for projected functions on the
constraint submanifold appears as a Lie bracket in the vector space }$%
\mathcal{O}$\textit{${}$ of equivalence classes }${E}_{f}$ \textit{of
functions in the phase space of quantum systems. The associative
star-product and the skew-symmetric Moyal bracket gifts ${}$}$\mathcal{O}$%
\textit{\ with a Poisson algebra structure.}

The mapping ${E}_{f}\leftrightarrow f_{t}$ induces further a Poisson algebra
structure on the vector space of projected functions$.$ The sum ${E}_{f}+{E}%
_{g}$ converts to $f_{t}+g_{t}$, the associative product ${E}_{f}{E}_{g}$
converts to $f_{t}\star g_{t}$, whereas the Dirac bracket $\{{E}_{f},{E}%
_{g}\}_{D}$ becomes the Moyal bracket $f_{t}\wedge g_{t}.$ It is clear that
these operations satisfy the Leibniz' law and the Jacoby identity and,
furthermore, keep the vector space of projected functions closed, as $%
(f_{t}+g_{t})_{t}=f_{t}+g_{t}$, $(f_{t}\star g_{t})_{t}=f_{t}\star g_{t}$,
and $(f_{t}\wedge h_{t})_{t}=f_{t}\wedge h_{t}$.

The skew-gradient projection of Eq.(\ref{PEV2}) gives an evolution equation
in the form involving projected functions only$.$ Eq.(\ref{PEV2}) is,
however, more convenient for applications, since it is valid for any
representative $f(\xi )$ of an equivalence class and does not presuppose
restrictions for $f(\xi ).$

In terms of the projected functions, the classical limit appears according
to equations 
\begin{eqnarray}
\lim_{\hbar \rightarrow 0}f_{t}\star g_{t} &=&f_{s}g_{s},  \label{CL1} \\
\lim_{\hbar \rightarrow 0}f_{t}\wedge g_{t} &=&\{f_{s},g_{s}\}=\{f,g\}_{D},
\label{CL2}
\end{eqnarray}
where the use is made of Eqs.(\ref{DB1}), (\ref{LIMIT}) and (\ref{FT2FS}).
It is worthwhile to notice that Eq.(\ref{DB1}) is valid on the constraint
submanifold, so the operation $f_{t}\wedge g_{t}$ defined for all functions
in the unconstrained phase space converts to the Dirac bracket on the
constraint submanifold only. Away from the constraint submanifold, functions
and, accordingly, the Dirac bracket do not make any physical sense. The
equivalence classes and the projected functions are the only objects
associated to physical observables.

In terms of the equivalence classes, the classical limit appears as 
\begin{eqnarray}
\lim_{\hbar \rightarrow 0}{E}_{f}{E}_{g} &=&\mathcal{E}_{f}\mathcal{E}_{g},
\label{CL3} \\
\lim_{\hbar \rightarrow 0}\{{E}_{f},{E}_{g}\}_{D} &=&\{\mathcal{E}_{f},%
\mathcal{E}_{g}\}_{D}.  \label{CL4}
\end{eqnarray}
The equivalence classes ${E}_{f}$ of the quantum systems represent the
quantum deformation of the equivalence classes $\mathcal{E}_{f}$ of the
corresponding classical systems. At $\hbar \rightarrow 0$, the star-product
quantization recovers the classical constrained dynamics.

To illustrate calculation of the quantum corrections to the Dirac bracket,
in Appendix C we derive the analytical expression for the $\hbar ^{2}$
correction.

%%%%%%%%%%%%%%%%%%%%%%%%%%%%%%%%%%%%%%%%%%%%%%%%%%%%%%%%%%%%%%%%%%%%%%%%

\section{Quantum spherical pendulum in the phase space}

\setcounter{equation}{0} 
%%%%%%%%%%%%%%%%%%%%%%%%%%%%%%%%%%%%%%%%%%%%%%%%%%%%%%%%%%%%%%%%%%%%%%%%

As an application we consider, in the phase space, the evolution of the
Wigner function of a mathematical pendulum on an $S^{n-1}$ sphere of a unit
radius in an $n$-dimensional Euclidean space with the coordinates $\phi
^{\alpha }$. 
%The $O(n)$ non-linear sigma model represents a field theory extension of the described model. 

At the classical level, there exists, within the generalized Hamiltonian
framework, two constraint functions 
\begin{equation}
{}\mathcal{G}^{1}=\ln \phi ,\;\;\;{}\mathcal{G}^{2}=\pi ^{\alpha }\phi
^{\alpha }  \label{OME}
\end{equation}
where $\phi =(\phi ^{\alpha }\phi ^{\alpha })^{1/2}$. The first constraint $%
{}\mathcal{G}^{1}=0$ implies that the particle stays on the sphere $\phi =1$%
, the second constraint ${}\mathcal{G}^{2}=0$ suggests that the radial
component of the momenta vanishes. The constraint functions constitute a
canonical pair (\ref{SB}). One can check that the projected canonical
variables $\xi _{s}(\xi )=(\phi _{s}^{\alpha },\pi _{s}^{\alpha })$, with 
\begin{eqnarray}
\phi _{s}^{\alpha } &=&\phi ^{\alpha }/\phi , \\
\pi _{s}^{\alpha } &=&\phi \pi ^{\alpha }-\phi ^{\alpha }\phi \pi /\phi ,
\end{eqnarray}
and the Hamiltonian projected onto the constraint submanifold (see \cite
{Krivoruchenko:2005wb}, Sect. 6), 
\begin{equation}
\mathcal{H}{}_{s}=\frac{1}{2}(\phi ^{2}\delta ^{\alpha \beta }-\phi ^{\alpha
}\phi ^{\beta })\pi ^{\alpha }\pi ^{\beta },  \label{HAMI}
\end{equation}
are identically in involution with ${}\mathcal{G}^{a}$, which is on a par
with Eqs.(\ref{CG}) and (\ref{XO}).

Since ${}\mathcal{G}^{2}$ is a second order polynomial with respect to the
canonical variables, the infinite power series in the Poisson operator in
Eq.(\ref{EA}) is truncated at $O(\hbar ^{0})$. Indeed, all the higher order
terms vanish, so that $\mathcal{G}{}^{1}\wedge {}\mathcal{G}^{2}=\{{}%
\mathcal{G}^{1},{}\mathcal{G}^{2}\}=1$. Assuming ${G}^{a}={}\mathcal{G}^{a}$%
, Eqs.(\ref{SBAS}) hold.

Similarly, Eqs.(\ref{CG3}) hold for $\xi _{t}(\xi )=\xi _{s}(\xi )$: ${}%
\mathcal{G}^{1}$ and $\phi _{s}^{\alpha }$ depend on $\phi ^{\alpha }$, so $%
\phi _{s}^{\alpha }\wedge {}\mathcal{G}^{1}=0$. $\pi _{s}^{\alpha }$ is a
first degree polynomial with respect to the canonical momenta. In the
expression $\pi _{s}^{\alpha }\wedge \mathcal{G}{}^{1}$, the operators ${}%
\mathcal{P}^{2k+1}$ act to the right on the coordinates only, and to the
left on the momenta. The power series terms entering Eq.(\ref{EA}) vanish
starting with $O(\hbar ^{2})$, so $\pi _{s}^{\alpha }\wedge {}\mathcal{G}%
^{1}=\{\pi _{s}^{\alpha },{}\mathcal{G}^{1}\}=0$ holds in virtue of Eqs.(\ref
{CG}). ${}\mathcal{G}^{2}$ is a second degree polynomial with respect to the
canonical variables. The infinite power series in Eq.(\ref{EA}) is truncated
at $O(\hbar ^{0})$ again. Due to Eqs.(\ref{CG}), we have ${}\mathcal{G}%
^{2}\wedge \phi _{s}^{\alpha }=\{{}\mathcal{G}^{2},\phi _{s}^{\alpha }\}=0$
and ${}\mathcal{G}^{2}\wedge \pi _{s}^{\alpha }=\{{}\mathcal{G}^{2},\pi
_{s}^{\alpha }\}=0$.

Finally, one has to check that ${}\mathcal{G}^{a}\wedge {}\mathcal{H}_{s}=0$%
. The first equation, for $a=1$, is valid since ${}\mathcal{G}^{1}$ depends
on $\phi ^{\alpha }$, while $\mathcal{H}_{s}$ depends on $\pi ^{\alpha }$
quadratically. The infinite power series in Eq.(\ref{EA}) is truncated at $%
O(\hbar ^{0})$, so the Moyal bracket can be replaced by the Poisson bracket.
Since $\{{}\mathcal{G}^{1},{}\mathcal{H}_{s}\}=0$, equation ${}\mathcal{G}%
^{1}\wedge {}\mathcal{H}_{s}=0$ holds. The second equation, for $a=2$, is
valid since ${}\mathcal{G}^{2}$ is a second degree polynomial. Consequently, 
$\mathcal{G}^{2}\wedge \mathcal{H}_{s}=\{ \mathcal{G}^{2}, \mathcal{H}_{s}
\}=0$.

The projected canonical variables $\xi _{s}(\xi )$, the constraint functions 
${}\mathcal{G}^{a}$, and the classical Hamiltonian $\mathcal{H}_{s}$
coincide with the projected canonical variables $\xi _{t}(\xi )$, the
constraint functions $G^{a}$, and the Hamiltonian $H_{t}$, respectively.

The evolution equation for the Wigner function has the form (\ref{PEV2})
where the sign of the right side should be changed. The power series
expansion of the Moyal bracket over the Poisson operator ${}$ is truncated
at $O(\hbar ^{2})$, since the Hamiltonian (\ref{HAMI}) is a fourth degree
polynomial of the canonical variables, so we obtain 
\begin{equation}
\frac{\partial }{\partial t}W=-\{W,{}\mathcal{H}_{s}\}+\frac{\hbar ^{2}}{8}(%
\frac{\partial ^{3}W}{\partial \phi ^{\alpha }\partial \phi ^{\beta
}\partial \pi ^{\gamma }}(2\delta ^{\alpha \beta }\phi ^{\gamma }-\delta
^{\alpha \gamma }\phi ^{\beta }-\delta ^{\beta \gamma }\phi ^{\alpha })-%
\frac{\partial ^{3}W}{\partial \pi ^{\alpha }\partial \pi ^{\beta }\partial
\phi ^{\gamma }}(2\delta ^{\alpha \beta }\pi ^{\gamma }-\delta ^{\alpha
\gamma }\pi ^{\beta }-\delta ^{\beta \gamma }\pi ^{\alpha })).  \label{PEV3}
\end{equation}
The first term in the right side is of the classical origin, while the
second term represents a quantum correction to the classical Liouville
equation and there are no other quantum corrections. Given $W(\xi ,0)$ in
the unconstrained phase space, $W(\xi ,t)$ can be found by solving the
partial differential equation (\ref{PEV3}).

%%%%%%%%%
\begin{table}[tbp]
\caption{
Skew-symmetric multiplication operations are listed for functions (second column) and equivalence classes of functions (third column) in the phase space of classical systems (first row) and in the phase space of quantum systems (second row).
}
\label{lab2}
\begin{center}
\begin{tabular}{|l|c|c|}
\hline
Systems: & unconstrained & constrained \\ \hline
classical & $\{f,g\} $ & $\{f,g\}_{D}$ \\ \hline
quantum & $f\wedge g$ & $f_{t}\wedge g_{t} $ \\ \hline
\end{tabular}
\end{center}
\par
\vspace{-2mm}
\end{table}
%%%%%%%%%%%

%%%%%%%%%%%%%%%%%%%%%%%%%%%%%%%%%%%%%%%%%%%%%%%%%%%%%%%%%%%%%%%%%%%%%%%%

\section{Conclusion}

\setcounter{equation}{0} 
%%%%%%%%%%%%%%%%%%%%%%%%%%%%%%%%%%%%%%%%%%%%%%%%%%%%%%%%%%%%%%%%%%%%%%%%

%{\bf 6. Conclusion.} 
Real functions in the unconstrained phase space of second-class constraints
systems split into equivalence classes corresponding to different physical
observables. The quantum observables are described by equivalence classes of
operators acting on the Hilbert space of states. The Weyl's association rule
extends the equivalence relations to functions in the phase space of quantum
systems.

The Dirac bracket can be calculated as the Poisson bracket between functions
projected onto the constraint submanifold using the phase flows generated by
the constraint functions. The quantum deformation of the Dirac bracket is
the Moyal bracket calculated for functions projected onto the constraint
submanifold of the phase space.

The skew-symmetric multiplication operations are synthesized in Table I.

\textit{The operation $f_{t}\wedge g_{t} $ designates the quantum
deformation of the Dirac bracket consistent with the canonical quantization
of constraint systems.}

As an application of the general formalism, in Sect. VI we derived with the
use of the operation $f_{t}\wedge g_{t}$ an evolution equation for the
Wigner function of an $n-1$-dimensional spherical pendulum, which represents
a mechanical counterpart of the $O(n)$ non-linear sigma model.

Our final conclusion states that quantum dynamics for constraint systems can
be formulated in the unconstrained phase space provided symplectic basis for
constraint functions exists globally.

\begin{acknowledgements}
This work is supported by DFG grant No. 436 RUS 113/721/0-1, RFBR grant No.
03-02-04004, and the European Graduate School Basel-T\"{u}bingen. 
\end{acknowledgements}

%%%%%%%%%%%%%%%%%%%%%%%%%%%%%%%%%%%%%%%%%%%%%%%%%%%%%%%%%%%%%%%%%%%%%%%%
\begin{appendix}
%%%%%%%%%%%%%%%%%%%%%%%%%%%%%%%%%%%%%%%%%%%%%%%%%%%%%%%%%%%%%%%%%%%%%%%%%%%%%%%%%%
%%%%%%%%%%%%%%%%%%%%%%%%%%%%%%%%%%%%%%%%%%%%%%%%%%%%%%%%%%%%%%%%%%%%%%%%%%%%%%%%%%

\section{Properties of the skew-gradient projection}

\setcounter{equation}{0} 
%%%%%%%%%%%%%%%%%%%%%%%%%%%%%%%%%%%%%%%%%%%%%%%%%%%%%%%%%%%%%%%%%%%%%%%%%%%%%%%%%%
%%%%%%%%%%%%%%%%%%%%%%%%%%%%%%%%%%%%%%%%%%%%%%%%%%%%%%%%%%%%%%%%%%%%%%%%%%%%%%%%%%

Let us apply Eq.(\ref{FSG}) to a function $f(\xi)$ expanded in a power
series in $\xi$: 
\begin{eqnarray}
f_{s}(\xi)&=& \sum_{k=0}^{\infty }\frac{1}{k!}\underbrace{\{...\{\{}%
_{k}f(\xi ),\mathcal{G}^{a_{1}}\},\mathcal{G}^{a_{2}}\},...\mathcal{G}^{a_{k}}\}
\mathcal{G}_{a_{1}}\mathcal{G}_{a_{2}}...\mathcal{G}_{a_{k}}  \nonumber \\
&\stackrel{2}{=}& \sum_{p=0}^{\infty }\frac{1}{p!}\frac{\partial ^{p}f(0)}{%
\partial \xi ^{i_{1}}...\partial \xi ^{i_{p}}}\sum_{k=0}^{\infty } \frac{1}{%
k!}\underbrace{\{...\{\{}_{k}\xi ^{i_{1}}...\xi
^{i_{p}},\mathcal{G}^{a_{1}}\},\mathcal{G}^{a_{2}}\},...\mathcal{G}^{a_{k}}\}
\mathcal{G}_{a_{1}}\mathcal{G}_{a_{2}}...\mathcal{G}_{a_{k}}  \nonumber \\
&\stackrel{3}{=}&\sum_{p=0}^{\infty }\frac{1}{p!}\frac{\partial ^{p}f(0)}{%
\partial \xi ^{i_{1}}...\partial \xi ^{i_{p}}}\sum_{k=0}^{\infty } \frac{1}{%
k!}\sum_{k_{1}+...+k_{p}=k}\frac{k!}{k_{1}!...k_{p}!}\underbrace{\{...\{\{}%
_{k_{1}}\xi ^{i_{1}},\mathcal{G}^{a_{1}}\},\mathcal{G}^{a_{2}}\},...\mathcal{G}^{a_{k_{1}}}\}  \nonumber
\\
&&...\underbrace{\{...\{\{}_{k_{p}}\xi ^{i_{p}},\mathcal{G}^{a_{k - k_{p} + 1}}\},
\mathcal{G}^{a_{k - k_{p} + 2}}\},...\mathcal{G}^{a_{k}}\} \mathcal{G}_{a_{1}}\mathcal{G}_{a_{2}}...\mathcal{G}_{a_{k}} 
\nonumber \\
&\stackrel{4}{=}&\sum_{p=0}^{\infty }\frac{1}{p!}\frac{\partial ^{p}f(0)}{%
\partial \xi ^{i_{1}}...\partial \xi ^{i_{p}}} \sum_{k_{1}\geq
0,...,k_{p}\geq 0}\frac{1}{k_{1}!...k_{p}!}\underbrace{\{...\{\{}_{k_{1}}\xi
^{i_{1}},\mathcal{G}^{a_{1}}\},\mathcal{G}^{a_{2}}\},...\mathcal{G}^{a_{k_{1}}}\}  \nonumber \\
&&...\underbrace{\{...\{\{}_{k_{p}}\xi ^{i_{p}},\mathcal{G}^{a_{k - k_{p} + 1}}\},
\mathcal{G}^{a_{k - k_{p} + 2}}\},...\mathcal{G}^{a_{k}}\} \mathcal{G}_{a_{1}}\mathcal{G}_{a_{2}}...\mathcal{G}_{a_{k}} 
\nonumber \\
&\stackrel{5}{=}&\sum_{p=0}^{\infty }\frac{1}{p!}\frac{\partial ^{p}f(0)}{%
\partial \xi ^{i_{1}}...\partial \xi ^{i_{p}}} \xi
^{i_{1}}_{s}(\xi)...\xi^{i_{p}}_{s}(\xi)  \nonumber \\
&=& f(\xi_{s}(\xi)).  \label{1LONG}
\end{eqnarray}
In this expression, the Taylor expansion is made first around $\xi=0$. To
reach the step 3 of Eq.(\ref{1LONG}), the multibinomial formula for
calculating the Poisson brackets of a product of $p$ functions is used.
Going from 3 to 4, the restriction $k_{1}+...+k_{p}=k$ is removed. The
summation over $k_{1},...,k_{p}$ becomes thereby independent. To achieve the
step 5 of Eq.(\ref{1LONG}), the summations over $k_{1},...,k_{p}$ are
performed, which turns $\xi$ into $\xi_{s}(\xi)$. The final result is given
by Eq.(\ref{FSSF}).

The classical counterpart of Eq.(\ref{POWE}) follows straightforwardly from
Eq.(\ref{FSSF}): 
\begin{equation}
(f_{1}...f_{p})_{s}(\xi )=f_{1s}(\xi )...f_{ps}(\xi ).  \label{MULTPR}
\end{equation}
In general, operators do not obey this property. However, under the
circumstances specified by Eq.(\ref{POWE}), this is still valid. Eq.(\ref
{MULTPR}) implies that $2^{p}$ products $\tilde{f}_{1}(\xi )...\tilde{f}%
_{p}(\xi )$ where $\tilde{f}_{i}(\xi )=f_{i}(\xi )$ or $\tilde{f}_{i}(\xi
)=f_{is}(\xi )$ for $1\leq i\leq p$ belong to the same equivalence class$.$
Eq.(\ref{MULTPR}) shows that class of the projected functions is closed
under the pointwise product.

Note other useful relation 
\begin{equation}
\{f,g_{s}\}_{s}=\{f_{s},g_{s}\}.  \label{A90}
\end{equation}
To check it, we apply Eq.(\ref{FSG}) for $\{f,g_{s}\}$ and use the Jacoby
identity. Since $g_{s}$ is identically in involution with $\mathcal{G}^{a}$,
one gets 
\begin{eqnarray*}
\{f,g_{s}\}_{s} &=&\sum_{k=0}^{\infty }\frac{1}{k!}\{\{...\{\{f,\mathcal{G}%
^{a_{1}}\},\mathcal{G}^{a_{2}}\},...\mathcal{G}^{a_{k}}\},g_{s}\}\mathcal{G}%
_{a_{1}}\mathcal{G}_{a_{2}}...\mathcal{G}_{a_{k}} \\
&=&\sum_{k=0}^{\infty }\frac{1}{k!}\{\{...\{\{f,\mathcal{G}^{a_{1}}\},%
\mathcal{G}^{a_{2}}\},...\mathcal{G}^{a_{k}}\}\mathcal{G}_{a_{1}}\mathcal{G}%
_{a_{2}}...\mathcal{G}_{a_{k}},g_{s}\} \\
&=&\{f_{s},g_{s}\}.
\end{eqnarray*}
In general, however, $\{f,g\}_{s}\neq \{f_{s},g_{s}\}.$ To see this, one can
set $f=\mathcal{G}^{a}$ and $g=\mathcal{G}^{b}$.

The classical analogue of Eq.(\ref{PROJCOM}) is given by 
\begin{equation}
(\{f,g\}_{D})_{s}=\{f_{s},g_{s}\}_{D}.  \label{A2}
\end{equation}
This equation is a consequence of Eqs.(\ref{DB1}) and, in particular, of the
fact that the Leibniz' law applies to the Dirac bracket, 
\begin{equation}
\{\{f,g\}_{D},\mathcal{G}^{a}\}=\{\{f,\mathcal{G}^{a}\},g\}_{D}+\{f,\{g,%
\mathcal{G}^{a}\}\}_{D}.  \label{A8}
\end{equation}
Using (\ref{A3}), we get 
\begin{eqnarray}
(\{f,g\}_{D})_{s} &=&\sum_{k=0}^{\infty }\frac{1}{k!}\{...\{\{\{f,g\}_{D},%
\mathcal{G}^{a_{1}}\},\mathcal{G}^{a_{2}}\},...\mathcal{G}^{a_{k}}\}\mathcal{%
G}_{a_{1}}\mathcal{G}_{a_{2}}...\mathcal{G}_{a_{k}}  \nonumber \\
&=&\sum_{k=0}^{\infty }\frac{1}{k!}\sum_{k_{1}+k_{2}=k}\frac{k!}{k_{1}!k_{2}!%
}\{\underbrace{\{...\{\{}_{k_{1}}f,\mathcal{G}^{a_{1}}\},\mathcal{G}%
^{a_{2}}\},...\mathcal{G}^{a_{k_{1}}}\},\underbrace{\{...\{\{}_{k_{2}}g,%
\mathcal{G}^{a_{k_{1}+1}}\},\mathcal{G}^{a_{k_{1}+2}}\},...\mathcal{G}%
^{a_{k}}\}\}_{D}\mathcal{G}_{a_{1}}\mathcal{G}_{a_{2}}...\mathcal{G}_{a_{k}}.
\nonumber
\end{eqnarray}
Taking into account $\{,\mathcal{G}^{a}\}_{D}=0$, one can place the last $k$
constraint functions inside of the Dirac bracket. The summation over $k$ can
be removed and we obtain 
\begin{eqnarray*}
&&\sum_{k_{1}\geq 0,k_{2}\geq 0}\frac{1}{k_{1}!k_{2}!}
\{\underbrace{\{...\{\{}_{k_{1}}f,\mathcal{G}^{a_{1}}\},\mathcal{G}^{a_{2}}\},...\mathcal{G}%
^{a_{k_{1}}}\}\mathcal{G}_{a_{1}}\mathcal{G}_{a_{2}}...\mathcal{G}%
_{a_{k_{1}}},\underbrace{\{...\{\{}_{k_{2}}g,\mathcal{G}^{a_{k_{1}+1}}\},%
\mathcal{G}^{a_{k_{1}+2}}\},...\mathcal{G}^{a_{k}}\}\mathcal{G}_{a_{k_{1}+1}}%
\mathcal{G}_{a_{k_{1}+2}}...\mathcal{G}_{a_{k}}\}_{D} \\
&=&\{f_{s},g_{s}\}_{D}.
\end{eqnarray*}
On the constraint submanifold, according to Eq.(\ref{DB1}), the subscript
can be omitted. In the unconstrained phase space, according to Eqs.(\ref
{DBNA}) and the fact that $f_{s}$ and $g_{s}$ are identically in involution
with the constraint functions, the subscript can be omitted also. Eq.(\ref
{A2}) shows that class of the projected functions is closed under the Dirac
bracket.

%%%%%%%%%%%%%%%%%%%%%%%%%%%%%%%%%%%%%%%%%%%%%%%%%%%%%%%%%%%%%%%%%%%%%%%%

\section{Properties of the operator function ${\tilde {\frak{B}}}(\eta)$}

\setcounter{equation}{0} 
%%%%%%%%%%%%%%%%%%%%%%%%%%%%%%%%%%%%%%%%%%%%%%%%%%%%%%%%%%%%%%%%%%%%%%%%

In order to check Eqs.(\ref{PROP}) - (\ref{PROP6}), it is useful to derive
first the similar properties for ${\tilde {\frak{B}}}(\eta)$: 
\begin{eqnarray}
\frak{\tilde{B}}(\eta )^{+} &=&\frak{\tilde{B}}(-\eta ),  \label{PT0} \\
Tr[\frak{\tilde{B}}(\eta )] &=&(2\pi \hbar )^{n}\delta ^{2n}(\eta ),
\label{PT1} \\
\frak{\tilde{B}}(0) &=&\frak{1},  \label{PT2} \\
\int \frac{d^{2n}\eta }{(2\pi \hbar )^{n}}\frak{\tilde{B}}(\eta )Tr[\frak{%
\tilde{B}}(-\eta )\frak{f}] &=&\frak{f}\ ,  \label{PT3} \\
Tr[\frak{\tilde{B}}(\eta )\frak{\tilde{B}}(-\eta ^{\prime })] &=&(2\pi \hbar
)^{n}\delta ^{2n}(\eta -\eta ^{\prime }),  \label{PT4} \\
\frak{\tilde{B}}(\eta )\frak{\tilde{B}}(-\eta ^{\prime }) &=&\frak{\tilde{B}}%
(\eta -\eta ^{\prime })\exp (-\frac{i}{2\hbar }\eta _{k}\eta _{l}^{\prime
}{I}^{kl}).  \label{PT5}
\end{eqnarray}
Eqs.(\ref{PT0}) and (\ref{PT2}) are conspicuous. Eq.(\ref{PT5}) can be
obtained using the identity $e^{A+B} = e^A e^B e^{-\frac{1}{2}[A,B]}$ that
holds for operators $A$ and $B$ whose commutator is a c-number. Eqs.(\ref
{PT1}) and (\ref{PT4}) can be derived by taking into account the explicit
form of the matrix elements of ${\tilde {\frak{B}}}(\eta)$, 
\begin{equation}
<\phi_{1}|{\tilde {\frak{B}}}(\eta)|\phi_{2}> = \delta^{n}(\phi^{\gamma}_{1}
- \phi^{\gamma}_{2} + \eta_{n + \gamma}) \exp(\frac{i}{2 \hbar}\sum_{\alpha
=1}^{n}\eta_{\alpha}(\phi^{\alpha}_{1} + \phi^{\alpha}_{2})),  \label{MATE}
\end{equation}
which can be obtained with the help of equation 
\[
\exp(\frac{i}{\hbar}\sum_{\alpha =1}^{n}\eta_{n+\alpha}\frak{x}%
^{n+\alpha})|\phi^{\gamma}> = |\phi^{\gamma} - \eta_{n+\gamma}>. 
\]
Using Eq.(\ref{MATE}) one gets 
\begin{equation}
\int \frac{d^{2n}\eta }{(2\pi \hbar )^{n}}<\phi_{1}|\frak{\tilde{B}}(\eta
)|\phi_{2}> <\phi_{3}|\frak{\tilde{B}}(-\eta )|\phi_{4}> =
\delta^{n}(\phi_{1}^{\alpha} -
\phi_{4}^{\alpha})\delta^{n}(\phi_{2}^{\alpha} - \phi_{3}^{\alpha}).
\label{COMPLETE}
\end{equation}
Since Eq.(\ref{PT3}) is valid for any operator, Eqs.(\ref{PT3}) and (\ref
{COMPLETE}) are equivalent.

Applying the Fourier transform to $\frak{\tilde{B}}(\eta )$, entering Eqs.(%
\ref{PT0}) - (\ref{PT5}), one gets Eqs.(\ref{PROP}) - (\ref{PROP6}).

Let us multiply Eq.(\ref{PROP3}) by $\frak{f}\frak{g}$, take the trace, and
use Eq.(\ref{GR}). As a consequence, we obtain 
\begin{equation}
Tr[\frak{f}\frak{g}] = \int \frac{d^{2n}\xi }{(2\pi \hbar )^{n}} f(\xi )
\circ g(\xi ).  \label{TRACE}
\end{equation}
%%%%%%%%%%%%%%%%%%%%%%%%%%%%%%%%%%%%%%%%%%%%%%%%%%%%%%%%%%%%%%
The symmetrized star-product can be replaced with the pointwise product.
Indeed, 
\begin{eqnarray}
f(\xi )\circ g(\xi ) &=&\sum_{p=0}^{\infty }\frac{(-1)^{p}}{(2p)!}\left(%
\frac{\hbar}{2}\right)^{2p}{I}^{i_{1}j_{1}}...{I}^{i_{2p}j_{2p}} \frac{%
\partial ^{2p}f(\xi )}{\partial \xi ^{i_{1}}...\partial \xi ^{i_{2p}}}\frac{%
\partial ^{2p}g(\xi )}{\partial \xi ^{j_{1}}...\partial \xi ^{j_{2p}}} \nonumber \\
&=&f(\xi )g(\xi ) + \sum_{p=1}^{\infty }\frac{(-1)^{p}}{(2p)!}\left(\frac{%
\hbar}{2}\right)^{2p}{I}^{i_{1}j_{1}}...{I}^{i_{2p}j_{2p}} \frac{\partial ^{2p}%
}{\partial \xi ^{i_{1}}...\partial \xi ^{i_{2p}}}\left( f(\xi )\frac{%
\partial ^{2p}g(\xi )}{\partial \xi ^{j_{1}}...\partial \xi ^{j_{2p}}}%
\right). \label{FORFAYA}
\end{eqnarray}
The quantum corrections $O(\hbar^{2p})$ which could make the variance
disappear, since they represent full derivatives and contribute to the
surface integral only. By multiplying Eq.(\ref{PROP4}) with $\frak{g}$ and
taking the trace, one arrives at the same conclusion.

%%%%%%%%%%%%%%%%%%%%%%%%%%%%%%%%%%%%%%%%%%%%%%%%%%%%%%%%%%%%%%%%%%%%%%%%%%%%%
\section{Lowest order quantum correction to the Dirac bracket}
%%%%%%%%%%%%%%%%%%%%%%%%%%%%%%%%%%%%%%%%%%%%%%%%%%%%%%%%%%%%%%%%%%%%%%%%%%%%%

The series expansion over the Planck's constant is straightforward, so we
restrict ourselves with the lowest order correction $O($ $\hbar ^{2})$.

The quantum deformation of the Dirac bracket is associated to the operation 
\begin{equation}
g_{t}(\xi )\wedge f_{t}(\xi )=(g(\xi )\wedge f_{t}(\xi ))_{t}.  \label{C1}
\end{equation}
Using Eq.(\ref{FT}), one can write 
\begin{eqnarray}
g_{t}(\xi )\wedge f_{t}(\xi ) &=&g(\xi )\wedge f_{t}(\xi )-\frac{\hbar ^{2}}{%
8}\{\{g(\xi ),f_{s}(\xi )\},{G}^{a_{1}}\}\mathcal{P}^{2}{G}_{a_{1}} 
\nonumber \\
&&-\frac{\hbar ^{2}}{8}\frac{1}{2!}\{\{\{g(\xi ),f_{s}(\xi )\},{G}^{a_{1}}\},%
{G}^{a_{2}}\}{G}_{a_{1}}\mathcal{P}^{2}{G}_{a_{2}}  \nonumber \\
&&-\frac{\hbar ^{2}}{8}\frac{1}{3!}\{\{\{\{g(\xi ),f_{s}(\xi )\},{G}%
^{a_{1}}\},{G}^{a_{2}}\},{G}^{a_{3}}\}{G}_{a_{1}}{G}_{a_{2}}\mathcal{P}^{2}{G%
}_{a_{3}}+O(\hbar ^{4}).  \label{C2}
\end{eqnarray}
In terms $O($ $\hbar ^{2}),$ the Moyal bracket is replaced with the Poisson
bracket and $f_{t}(\xi )$ is replaced with $f_{s}(\xi )$. The arguments
similar to that used for the series expansion (\ref{FT}) allow to truncate
the series expansion over ${G}_{a},$ entering $f_{s}$: 
\begin{eqnarray}
g_{t}(\xi )\wedge f_{t}(\xi ) &=&g(\xi )\wedge f_{t}(\xi )  \nonumber \\
&&-\frac{\hbar ^{2}}{8}\{\{g(\xi ),{G}^{a_{1}}\},f(\xi )+\{f(\xi ),{G}%
^{a_{2}}\}{G}_{a_{2}}+\frac{1}{2!}\{\{f(\xi ),{G}^{a_{2}}\},{G}^{a_{3}}\}{G}%
_{a_{2}}{G}_{a_{3}}  \nonumber \\
&&+\frac{1}{3!}\{\{\{f(\xi ),{G}^{a_{2}}\},{G}^{a_{3}}\}{G}^{a_{4}}\}
{G}_{a_{2}}{G}_{a_{3}}{G}_{a_{4}}\}\mathcal{P}^{2}{G}_{a_{1}}  \nonumber \\
&&-\frac{\hbar ^{2}}{8}\frac{1}{2!}\{\{\{g(\xi ),{G}^{a_{1}}\},{G}%
^{a_{2}}\},f(\xi )+\{f(\xi ),{G}^{a_{3}}\}{G}_{a_{3}}+\frac{1}{2!}\{\{f(\xi
),{G}^{a_{3}}\},{G}^{a_{4}}\}{G}_{a_{3}}{G}_{a_{4}}\}{G}_{a_{1}}\mathcal{P}%
^{2}{G}_{a_{2}}  \nonumber \\
&&-\frac{\hbar ^{2}}{8}\frac{1}{3!}\{\{\{\{g(\xi ),{G}^{a_{1}}\},{G}%
^{a_{2}}\},{G}^{a_{3}}\},f(\xi )+\{f(\xi ),{G}^{a_{4}}\}{G}_{a_{4}}\}{G}%
_{a_{1}}{G}_{a_{2}}\mathcal{P}^{2}{G}_{a_{3}}+O(\hbar ^{4}).  \label{FI}
\end{eqnarray}
Let us turn to the first term of Eq.(\ref{FI}). We have 
\begin{eqnarray}
g(\xi )\wedge f_{t}(\xi ) &=&(g(\xi )\wedge f(\xi ))_{t}  \nonumber \\
&&+\sum_{k=1}^{\infty }\frac{1}{k!}\sum_{l=1}^{k}(...(((...((f(\xi )\wedge {G}%
^{a_{1}})\wedge {G}^{a_{2}})...\wedge {G}^{a_{l-1}})\wedge (g(\xi )\wedge {G}%
^{a_{l}}))\wedge {G}^{a_{l+1}})...\wedge {G}^{a_{k}})  \nonumber \\
&&\circ {G}_{a_{1}}\circ {G}_{a_{2}}...\circ {G}_{a_{k}}+  \nonumber \\
&&+\sum_{k=1}^{\infty }\frac{1}{k!}\sum_{l=1}^{k}(...((f(\xi )\wedge {G}%
^{a_{1}})\wedge {G}^{a_{2}})...\wedge {G}^{a_{k}})  \nonumber \\
&&\circ {G}_{a_{1}}\circ {G}_{a_{2}}...\circ {G}_{a_{l-1}}\circ (g(\xi
)\wedge {G}_{a_{l}})\circ {G}_{a_{l+1}}...\circ {G}_{a_{k}}.  \label{1111}
\end{eqnarray}
The first term can be expanded using Eq.(\ref{FT}). In the second term, the
Moyal bracket can be replaced with the Poisson bracket. The symmetrized
star-product $\circ $ is treated as in Eq.(\ref{FT}). The series over $k$ to
order $O(\hbar ^{2})$ is truncated at $k=3$ again. The third term is
truncated at $k=4.$ The Dirac bracket originates, to the zeroth order in $%
\hbar ,$ from the first term with $k=0$ and the third term with $k=1$ in Eq.(%
\ref{1111}). The result takes the form 
\begin{eqnarray}
g(\xi )\wedge f_{t}(\xi ) &=&\{g(\xi ),f(\xi )\}_{D} \nonumber \\
&&- \frac{\hbar^2}{24}g(\xi)\mathcal{P}^3f(\xi)  \nonumber \\
&&-\frac{\hbar ^{2}}{8}\{\{f(\xi ),g(\xi )\},{G}^{a_{1}}\}\mathcal{P}^{2}{G}%
_{a_{1}}-\frac{\hbar ^{2}}{8}\frac{1}{2!}\{\{\{f(\xi ),g(\xi )\},{G}%
^{a_{1}}\},{G}^{a_{2}}\}{G}_{a_{1}}\mathcal{P}^{2}{G}_{a_{2}}  \nonumber \\
&&-\frac{\hbar ^{2}}{8}\frac{1}{3!}\{\{\{\{f(\xi ),g(\xi )\},{G}^{a_{1}}\},{G%
}^{a_{2}}\},{G}^{a_{3}}\}{G}_{a_{1}}{G}_{a_{2}}\mathcal{P}^{2}{G}_{a_{3}} 
\nonumber \\
&&-\frac{\hbar ^{2}}{24}(f(\xi )\mathcal{P}^{3}{G}^{a_{1}})\{g(\xi ),{G}%
_{a_{1}}\}-\frac{\hbar ^{2}}{24}\{f(\xi ),{G}^{a_{1}}\}(g(\xi )\mathcal{P}%
^{3}{G}_{a_{1}})-\frac{\hbar ^{2}}{8}\{f(\xi ),{G}^{a_{1}}\}\mathcal{P}%
^{2}\{g(\xi ),{G}_{a_{1}}\}  \nonumber \\
&&-\frac{\hbar ^{2}}{8}\{f(\xi ),\{g(\xi ),{G}^{a_{1}}\}\}\mathcal{P}^{2}{G}%
_{a_{1}}  \nonumber \\
&&-\frac{\hbar ^{2}}{8}\frac{1}{2!}\{\{f(\xi ),\{g(\xi ),{G}^{a_{1}}\}\},{G}%
^{a_{2}}\}{G}_{a_{1}}\mathcal{P}^{2}{G}_{a_{2}}  \nonumber \\
&&-\frac{\hbar ^{2}}{8}\frac{1}{2!}\{\{f(\xi ),{G}^{a_{1}}\},\{g(\xi ),{G}%
^{a_{2}}\}\}{G}_{a_{1}}\mathcal{P}^{2}{G}_{a_{2}}  \nonumber \\
&&-\frac{\hbar ^{2}}{8}\frac{1}{3!}\{\{\{f(\xi ),\{g(\xi ),{G}^{a_{1}}\}\},{G%
}^{a_{2}}\},{G}^{a_{3}}\}{G}_{a_{1}}{G}_{a_{2}}\mathcal{P}^{2}{G}_{a_{3}} 
\nonumber \\
&&-\frac{\hbar ^{2}}{8}\frac{1}{3!}\{\{\{f(\xi ),{G}^{a_{1}}\},\{g(\xi ),{G}%
^{a_{2}}\}\},{G}^{a_{3}}\}{G}_{a_{1}}{G}_{a_{2}}\mathcal{P}^{2}{G}_{a_{3}} 
\nonumber \\
&&-\frac{\hbar ^{2}}{8}\frac{1}{3!}\{\{\{f(\xi ),{G}^{a_{1}}\},{G}%
^{a_{2}}\},\{g(\xi ),{G}^{a_{3}}\}\}{G}_{a_{1}}{G}_{a_{2}}\mathcal{P}^{2}{G}%
_{a_{3}}  \nonumber \\
&&-\frac{\hbar ^{2}}{8}\frac{1}{2!}(\{\{f(\xi ),{G}^{a_{1}}\},{G}^{a_{2}}\}%
\mathcal{P}^{2}{G}_{a_{1}})\{g(\xi ),{G}_{a_{2}}\}  \nonumber \\
&&-\frac{\hbar ^{2}}{8}\frac{1}{2!}\{\{f(\xi ),{G}^{a_{1}}\},{G}^{a_{2}}\}{G}%
_{a_{1}}\mathcal{P}^{2}\{g(\xi ),{G}_{a_{2}}\}  \nonumber \\
&&-\frac{\hbar ^{2}}{8}\frac{1}{2!}\{\{f(\xi ),{G}^{a_{1}}\},{G}%
^{a_{2}}\}\{g(\xi ),{G}_{a_{1}}\}\mathcal{P}^{2}{G}_{a_{2}}  \nonumber \\
&&-\frac{\hbar ^{2}}{8}\frac{1}{3!}(\{\{\{f(\xi ),{G}^{a_{1}}\},{G}%
^{a_{2}}\},{G}^{a_{3}}\}{G}_{a_{1}}\mathcal{P}^{2}{G}_{a_{2}})\{g(\xi ),{G}%
_{a_{3}}\}  \nonumber \\
&&-\frac{\hbar ^{2}}{8}\frac{1}{3!}\{\{\{f(\xi ),{G}^{a_{1}}\},{G}^{a_{2}}\},%
{G}^{a_{3}}\}{G}_{a_{1}}{G}_{a_{2}}\mathcal{P}^{2}\{g(\xi ),{G}_{a_{3}}\} 
\nonumber \\
&&-\frac{\hbar ^{2}}{8}\frac{2}{3!}\{\{\{f(\xi ),{G}^{a_{1}}\},{G}^{a_{2}}\},%
{G}^{a_{3}}\}\{g(\xi ),{G}_{a_{1}}\}{G}_{a_{2}}\mathcal{P}^{2}{G}_{a_{3}} 
\nonumber \\
&&-\frac{\hbar ^{2}}{8}\frac{4}{4!}\{\{\{\{f(\xi ),{G}^{a_{1}}\},{G}%
^{a_{2}}\},{G}^{a_{3}}\},{G}^{a_{4}}\}\{g(\xi ),{G}_{a_{1}}\}{G}_{a_{2}}{G}%
_{a_{3}}\mathcal{P}^{2}{G}_{a_{4}}+O(\hbar ^{4})  \label{OE}
\end{eqnarray}
In the first line, we recover the Dirac bracket (\ref{DBNA}). The second term 
is the quantum correction to the Poisson bracket. The third
and fourth lines originate from the expansion of the first term in Eq.(\ref
{1111}) as prescribed by Eq.(\ref{FT}). The fifth line comes from the
expansion of the $k=1$ component of the third term in Eq.(\ref{1111}). The
lines 6 - 11 appear from the expansion of the second term in Eq.(\ref{1111}%
). The rest appears from the components $k>1$ of the third term in Eq.(\ref
{1111}). The round brackets unify terms within the action of the operation $%
\mathcal{P}^{2}$. If the round brackets are suppressed (in the most cases), $%
\mathcal{P}^{2}$ acts on all terms.

Combining Eqs.(\ref{FI}) and (\ref{OE}), we get the Dirac bracket on the
constraint submanifold to the order $O(\hbar ^{2})$.

\end{appendix}

%%%%%%%%%%%%%%%%%%%%%%%%%%%%%%%%%%%%%%%%%%%%%%%%%%%%%%%%%%%%%%%%%%%%%%%%

\end{document}